\documentstyle[multicol,pre,aps,epsfig]{revtex}

\begin{document}

\author{Mario Castro}
\address{
Grupo Interdisciplinar de Sistemas Complejos (GISC) \&  \\
Escuela T\'ecnica Superior de Ingenier\'{\i}a (ICAI), \\
Universidad Pontificia Comillas, E-28008 Madrid, Spain} 

\title{Phase-field approach to heterogeneous nucleation}
\maketitle

\begin{abstract}
We consider the problem of heterogeneous nucleation and growth. The system
is described by a phase field model in which the temperature is included
through thermal noise. We show that this phase field approach is
suitable to describe homogeneous as well as heterogeneous nucleation
starting from several general hypotheses. Thus we can investigate the influence
of grain boundaries, localized impurities, or any general kind of
imperfections in a systematic way.
We also put forward the applicability of our model to study
other physical situations such as island formation, amorphous crystallization,
or recrystallization.
\end{abstract}


\begin{multicols}{2}
\narrowtext

\section{Introduction}

Processes driven by nucleation and growth have attracted much attention
during past decades, from a fundamental point of view~\cite{Gunton} and 
for tailoring some technological applications. Some of them are: 
the recrystallization of deformed metals~\cite{Doherty}, controlling 
the nucleation and growth of islands on terraces
in order to get large scale arrays of nanostructures~\cite{Castellano}, or
the manufacturing of thin-film transistors which are the basic devices for some 
applications as solar cells~\cite{Bergmann}, random 
access static memories~\cite{Im}, or active matrix-addressed flat-panel 
displays~\cite{Uemoto}. 

In all the above-mentioned processes, a metastable phase decays into a 
stable one via a 
fluctuation which produces a critical cluster of atoms (for instance, a
critical island in the case of terrace growth, or a critical 
atom cluster in the case of crystallization). This transition is
called nucleation.  At a certain fixed temperature,
clusters with sizes greater than a critical one become stable
nuclei; otherwise they shrink and eventually vanish. Such a
critical size arises from the competition  between the surface tension
and the chemical potential difference between phases, yielding an energy 
barrier that has to be overcome to build up a critical nucleus. 
For the examples presented above, the system can be considered, under certain
conditions, two dimensional so it is straightforward to write the free energy 
of circular grain of radius $r$:
\begin{equation}
\Delta F(r)=2\pi r\sigma-\pi r^2\Delta\mu/\Omega,
\end{equation}
where, $\sigma$, $\Delta \mu$, and $\Omega$ are the surface tension, the
chemical potential difference between phases, and the mean volume occupied
by an atom, respectively.

Notwithstanding, in some practical situations this transformation is not 
perfectly homogeneous due to, for instance, the presence of physical
boundaries, such as terrace steps, the interplay between different kinds of
particles, or the appearance of impurities~\cite{Meso}, or even to 
some preexisting order embedded in the initial phase formed during its 
manufacturing~\cite{Olivares}.
The lack of uniformity not only catalyzes the transformation but affects 
the final size distribution of grains~\cite{us1,us2}.

Much work has been devoted to the study of homogeneous nucleation
in different contexts~\cite{Gunton,Grant,Roy,Rikvold}, and to 
the so-called site saturation in which nucleation takes place
just at the beginning of the transformation~\cite{Jou}, but only a few 
studies were devoted to
the intermediate situations in which nucleation is heterogeneous,
both in space and in time. The spatial extent of heterogeneity can be
measured using the fraction of transformed material (or more generically,
the volume fraction of the newly transformed phase) onto the stable 
phase which often obeys the 
Kolmogorov-Johnson-Mehl-Avrami (KJMA) equation~\cite{Kolmogorov}:
\begin{equation}
X(t)=1-\exp\left\{-\left[(t-\tau_i)/\tau_c\right]^m\right\},
\label{cryst_frac}
\end{equation}
where $\tau_i$ is the incubation time, $\tau_c$ is the characteristic 
transformation time,
and $m$ is an exponent which characterizes the degree of
heterogeneity of the system and its dimensionality~\cite{us1,us2}. 
For instance, for a two-dimensional system in the two limiting cases, 
homogeneous nucleation and site saturation, $m$ takes the values $3$ and $2$,
respectively.

There have been some approaches to the problem of heterogeneous 
nucleation in the past.  For instance, Karpov 
{\em et al.}~\cite{Karpov} generalized the homogeneous case by adding 
random contributions to the surface tension and the chemical potential, which 
affects the nucleation rate.  A similar {\em static} approach was 
followed by Liu~\cite{Liu}, who calculated the variation suffered by 
the surface tension and the chemical potential due to the presence of a circular 
impurity. Other authors such as Enomoto~\cite{Enomoto} or
Weinberg~\cite{Weinberg} considered a phenomenological time-dependent
nucleation rate to analyze heterogeneous nucleation. 
More recently, Castro and co-workers~\cite{us1,us2} have introduced a lattice 
model to determine both time-dependent and spatial effects of heterogeneities.

The main aim of this paper is analyze, both analytically and
numerically, the origin and effect of impurities or defects 
on grain nucleation and subsequent growth. The rest of the paper 
is organized as follows. In Sec.\ \ref{ii} we introduce a phase field model 
obtained from a functional free energy and report some results for 
homogeneous nucleation. In Sec.\ \ref{iii}, we include impurities 
through the boundary conditions of the main equation and calculate the 
effect of those impurities in nucleation rate and KJMA exponent. 
In Sec.\ \ref{iv}, we consider a more general case in which the disorder 
can be due to mechanisms which are different with the ones
presented in Sec.\ \ref{iii}. These mechanisms are introduced as quenched noise.   
Finally, we end the paper by summarizing the main 
results in Sec.\ V, focusing on the applicability of the equations to
different physical processes, and discussing further generalizations of the 
model.

\section{Evolution equation}
\label{ii}

Phase-field models have been widely studied in the last few years,
as an efficient computational tool to simulate some moving boundary 
problems which, in the so-called sharp interface limit (or sometimes
thin interface limit, see below), are physically 
equivalent~\cite{Jou,Kobayashi,Wheeler,Karma,Folch,Elder,Collins,Tapio}.
Among them, Jou and Lusk~\cite{Jou} studied 
homogeneous nucleation and site saturation using a one-field phase field
model. The main objection of their approach is the fact that the
critical clusters are created {\em ad hoc} so, on the one hand, the model 
explains the KJMA equation just by construction and, on the other hand, it cannot
explain the existence of an incubation time observed in the experiments. 
A similar approach was used by
Roy {\em et al.}~\cite{Roy} to study nucleation in a phase field model
with nonlocal interactions. Recently, more complex phase field
models have been proposed for similar systems: The so-called multi-phase-field 
models~\cite{Steinbach,Nestler}, in which 
every cluster appearing embedded in the metastable phase is described
by its own field and, on the other hand, those models in 
which the phase field is coupled with another field representing the 
orientation of crystalline planes~\cite{Kobayashi2,Lobkovsky}.
The main problem concerning the multiphase-field models is that the free
energy depends explicitly on the grain orientation (or phase).  This is
solved by the second ones but, besides this, what depends on 
the grain orientation is the grain-boundary velocity. Although this second 
kind of approach seems to be very promising, in both cases the symmetry under 
rotations of the grain crystalline planes is broken.

In this paper we are interested only in the overall dynamics of 
the nucleation process, thus, we simply generalize the model presented 
in Ref.\ ~\cite{Jou}, supplementing it with thermal noise to make explicit the 
temperature dependence of the system. 
To make clearer this generalization, we advance that the
noise term is the driving force for nucleation, so it is not needed to create 
artificially critical clusters every integration time step as in 
Ref.\ \cite{Jou}.

Let us introduce the main ingredients of the model.  
We define an order parameter $\phi$, which takes the value $-1$ in the
metastable phase and $+1$ in the stable phase. The grain boundary (which
separates both phases) is located at $\phi=0$~\cite{phi0}.
We also define a free-energy functional which takes into account the
grain-boundary energy and the chemical potential difference between
phases. Generically, we can define 
\begin{equation}
{\mathcal F}[\phi({\bf r},t)]=\int dxdz \Big(\frac{W^2}{2}|\nabla
\phi|^2+f(\phi)-\lambda g(\phi) \Big), 
\label{Gibbs_funcional}
\end{equation} 
where $f(\phi)$ and $g(\phi)$ are generic functions of 
the order parameter:
$f(\phi)$ is an even function of $\phi$ with local minima at $\pm 1$ and
$g(\phi)$ breaks the symmetry between phases.
As we will see below, $W$ is a typical length scale related to the surface
tension, and $\lambda$ is a dimensionless parameter proportional to the 
chemical potential difference between phases. 
We will assume that the system relaxes towards equilibrium according
to the following evolution equations~\cite{Hohenberg}:
\begin{eqnarray}
\tau\partial_t\phi&=&-\frac{\delta{\mathcal F}}{\delta\phi}+\theta
\Rightarrow\nonumber\\
\tau\partial_t\phi&=&W^2\nabla^2\phi-f_\phi(\phi)+\lambda g_\phi+\theta,
\label{pf}
\end{eqnarray}
where $\tau$ is the typical time scale at which the atoms from a phase
incorporate to the other; $f_\phi$ and $g_\phi$ 
denote the partial derivatives of $f$ and $g$ with respect 
to $\phi$.
Finally, $\theta({\bf r},t)$ is a Gaussian white noise which stands
for the thermal fluctuations of the system, with zero mean and 
correlations given by the fluctuation-dissipation theorem~\cite{Hohenberg}:
\begin{equation}
\big\langle \theta({\bf r},t)\theta({\bf r}^\prime,t^\prime)\big\rangle=
2\tau k_BT\delta({\bf r}-{\bf r}^\prime)\delta(t-t^\prime),
\label{fluc_disip}
\end{equation}
$T$ being the temperature at which the transformation takes place.

To be more specific, we choose
\begin{equation}
f(\phi)=-\phi^2/2+\phi^4/4,\label{fdef}
\end{equation}
The main advantage of this choice is that 
Eq.\ (\ref{pf}) admits a simple stationary solution given by:
\begin{equation}
\phi_0(z)=-\tanh\left(\frac{z}{\sqrt{2}W} \right),
\label{tanh}
\end{equation}
which represents a front of characteristic width $W$ placed at $z=0$. 
In the same way, we choose~\cite{other}:
\begin{equation}
g(\phi)= \phi-\frac{\phi^3}{3}.\label{gdef}
\end{equation}
The main reason to use Eq.\ (\ref{gdef}) instead of the traditional
one, $g(\phi)=\phi$, is that, in the first case,  $\mathcal F$ has local
minima at $\pm 1$ independently of the value of $\lambda$, otherwise, those
minima would be $\lambda$ dependent. 

Using Eq.\ (\ref{tanh}) 
we can make some considerations about the stability of a given fluctuation.
To compare with the classical nucleation theory~\cite{cnt}, 
let us consider  
the free energy difference between a system which is initially at the 
metastable phase, and a circular grain of radius $R$ given
approximately by $\phi_0(r-R)$. Thus it can be straightforwardly shown that,
in the thin interface limit~\cite{Karma}:
\begin{equation}
\Delta{\mathcal F}\equiv {\mathcal F}[\phi=\phi_0]-{\mathcal F}[\phi=-1]\simeq 
2\pi R\sigma_l-\pi R^2\Delta\mu/\Omega,
\label{thin_interface}
\end{equation}
where $\sigma_l=2\sqrt{2}W/3$ is the surface tension, and 
$\Delta\mu/\Omega=\lambda\delta g$, with $\delta g=g(+1)-g(-1)$.
This equation shows the competition between the gain arising from the reduction
of the grain perimeter, and that related to the increasing of its size. 
As we mentioned above, the critical radius arises from this competition.  
Thus if we take into account that $\exp[-\Delta{\mathcal F}(R)/k_BT]$ can
be interpreted as the barrier that has to be overcome to create a nucleus of 
radius $R$, then the critical radius is the one which
minimizes that barrier,  the one which maximizes 
$\Delta{\mathcal F}$. Hence
\begin{equation}
R^*=\frac{2\sqrt{2}W}{3\lambda\delta g},
\label{criticalR}
\end{equation}
and its corresponding critical free energy is given by
\begin{equation}
\Delta {\mathcal F}^*=\frac{8\pi W^2}{9\lambda\delta g}.
\label{criticalF}
\end{equation}

\begin{figure}[!ht]
\begin{center} \epsfxsize=7cm \epsffile{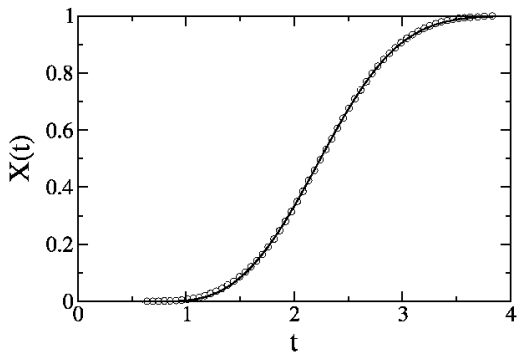} \end{center}
\caption{Transformed fraction $X(t)$ vs time for a $500\times
500$ system. ($\circ$) numerical integration of Eq.\ (\ref{pf}) 
with $W=1$, $\tau=0.15$, $\lambda=0.6$,  $\Delta x=1$,
$\Delta t=0.01$, and  $k_BT=0.1$, averaged over $100$ runs.
Solid line, KJMA fit with $\tau_{i}=0.75\pm 0.01$ and $\tau_c=1.68\pm0.01$.}
\label{test_GL_KJMA}
\end{figure}

Once we have established the connection between the physical system and
the model, let us numerically check its capability to reproduce the
KJMA equation (\ref{cryst_frac}). 
In Fig.\ \ref{test_GL_KJMA} we show the results from simulation using
the Euler integration scheme~\cite{Gardiner}, and the corresponding KJMA fit to the
fraction of transformed material, $X(t)$, taking $m=3$ 
(measured as the fraction of sites where $\phi>0$). 
We have reproduced this result for a wide range of parameters.
The first important result concluded from these simulations
is related to the nonzero
value of the incubation time $\tau_i$. This time is almost always present
in the experiments, and is related to the free-energy difference between
the well $\phi=-1$ and the maximum of the free-energy separating this
well with the one at $\phi=+1$. 
Similar results were obtained by Elder {\em et al.}~\cite{Elder_eu} in the 
context of eutectic growth.
Despite the good agreement between simulations and theory, 
it is important to stress that
the derivation of KJMA equation makes uses of some assumptions: infinite
system size, uniform nucleation, spherical particles, and constant 
growth rate. 
Some care must be taken in this respect.  Despite Eq.\ (\ref{pf}) provides circular 
grains, those grains do not grow at constant velocity. Actually, 
the grain radius is related to time through the following equation:
\begin{equation}
t=\frac{1}{V}\left[R-R_0+R^*\ln\left(\frac{R-R^*}{R_0-R^*}\right)\right],
\end{equation}
where $R_0$ is the initial grain radius and $V$ is the grain growth velocity which, 
in the thin interface limit, is given by
\begin{equation}
V=\frac{3W\lambda\delta g}{\sqrt{8}\tau}.
\end{equation}
Thus the velocity is constant just after a short time of order $R^*/V$. Hence
there are slight deviations of the numerical data from the KJMA equation up to
times of order $R^*/V$.
Besides this, the infinite size condition has not to be fulfilled necessarily. Actually,
the KJMA equation is still valid when the number of grains contributing to growth is large,
when the system size $L$, is much larger than the characteristic length scale
related to nucleation and growth, $l=(V/N)^{1/3}$, $N$ being the nucleation rate~\cite{Rikvold}.
Finally, the KJMA equation does not take into account interfacial effects that govern
the growth just before the grains meet each other, so there are also some differences
between the numerical data and the KJMA equation predictions, in this case, at the
later stages of the transformation. Fortunately, the mentioned
deviations from the KJMA equation at short and long times are quite small. Notwithstanding,  
it is quite convenient to perform the fitting of $X(t)$ between 
the $1\%$ and the $99\%$ to improve it
(see Ref.~\cite{Rikvold} for further details about the validity of KJMA equation). 

\begin{figure}[!ht]
\begin{center} \epsfxsize=9cm \epsffile{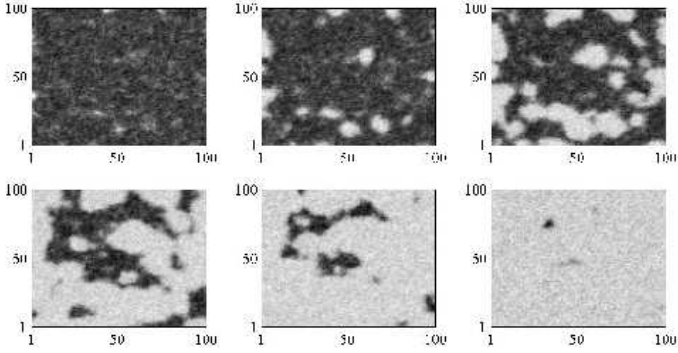} \end{center}
\caption{Numerical integration of Eq.\ (\ref{pf}) 
with $W=1$, $\tau=0.15$, $\lambda=0.6$,  $\Delta x=1$,
$\Delta t=0.01$, and  $k_BT=0.1$. From left to right, from top to bottom, 
corresponding times $1.0$, $1.5$, $2.0$, $2.5$, $3.0$, and $3.5$. 
The darker sites stand for $\phi\simeq -1$ and the brighter ones 
for $\phi\simeq +1$. It can be clearly seen how the initial metastable 
phase evolves into almost circular grains which nucleate and coalesce 
continuously.}
\label{evol_kjma}
\end{figure}

To point out the crucial role played by thermal noise in Eq.\ (\ref{pf}),
we show in Fig.\ \ref{evol_kjma} the spontaneous and continuous nucleation 
and growth of grains,  which are almost circular despite the 
underlying integration lattice is square. In other words, the model
captures all the essential ingredients of a first-order transition, not
only in terms of the free-energy difference between phases, but also in 
terms of the dynamical
path followed from the metastable phase to the stable one.

\section{Nucleation at defects}
\label{iii}

In the previous section we have demonstrated the validity of our model
to describe homogeneous nucleation. 
This section deals with nucleation in the neighborhood of some 
parts of the system which are structurally different, namely, 
at domain walls or at the surrounding of foreign particles.
Thus we will consider two kinds of defects: Walls and circular 
defects.  Our main assumption here is that the particles in the metastable phase 
do not interact with the defects, which is included in the model through the 
boundary condition
\begin{equation}
\frac{\partial\phi}{\partial n}=\nabla\phi\cdot{\bf{n}}|_b=0,
\label{bound_cond}
\end{equation}
where the subscript $b$ stand for boundary; and
$n$ is the normal coordinate to the defect boundary.

\begin{figure}[!ht]
\begin{center} \epsfxsize=9cm \epsffile{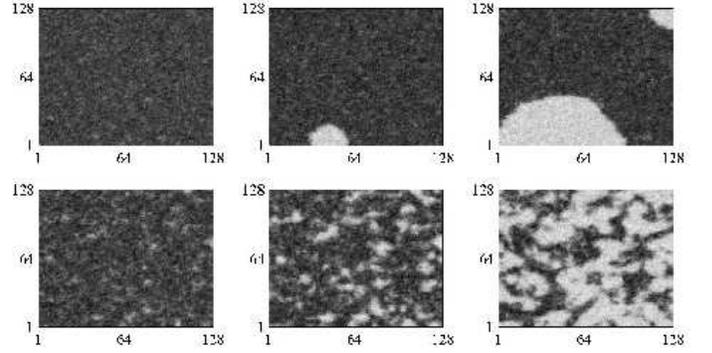} \end{center}
\caption{Numerical integration of Eq.\ (\ref{pf}) using boundary condition
(\ref{bound_cond}), with $W=1$, $\tau=0.1$,
$\Delta x=1$, $\Delta t=0.005$, and  $k_BT=0.15$. The top three figures 
correspond to $\lambda=0.35$ at times $1$, $5$, and $10$. The bottom
three figures correspond to $\lambda=0.6$ at times $0.50$, $0.75$, and $1.00$.} 
\label{impur_pared}
\end{figure}

In order to clarify the relevance of this boundary condition on nucleation
we have integrated Eq.\ (\ref{pf}) with the prescribed condition 
(\ref{bound_cond}).  Figure \ref{impur_pared} shows how nucleation is

enhanced at the walls. Actually, this is the most relevant
mechanism of nucleation when the chemical potential difference between
phases, $\Delta\mu\propto \lambda$, is small. 
The situation changes dramatically if we increase $\lambda$ (or if we
raise the temperature). This can
be better understood in the context of island formation in which $\lambda$
can be understood as the flux of particles arriving at the surface. In such
case, when the flux of particles is large, islands nucleate
everywhere in the sample due to the large probability of dimer 
formation~\cite{Castellano}, as can be seen in Fig.\ \ref{impur_pared}.

\begin{figure}[!ht]
\begin{center} \epsfxsize=9cm \epsffile{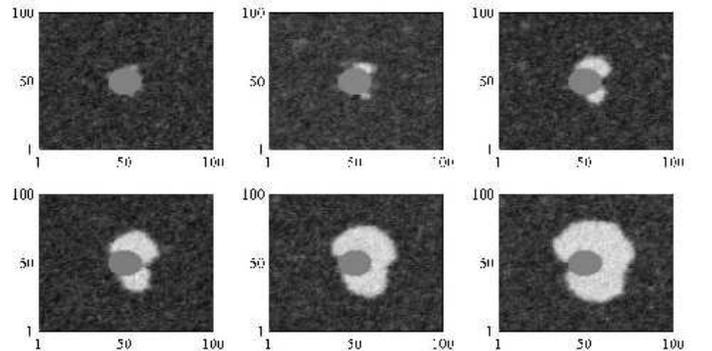} \end{center}
\caption{Numerical integration of Eq.\ (\ref{pf}) using boundary condition
(\ref{bound_cond}) over the circle perimeter, with $W=1$, $\tau=0.1$,
$\lambda=0.5$, $\Delta x=1$, $\Delta t=0.005$, and  $k_BT=0.1$.
From left to right and from top to bottom, corresponding to 
equally spaced times from $1$ to $6$.  } 
\label{impur_circ}
\end{figure}

\begin{figure}[!ht]
\begin{center} \epsfxsize=7cm \epsffile{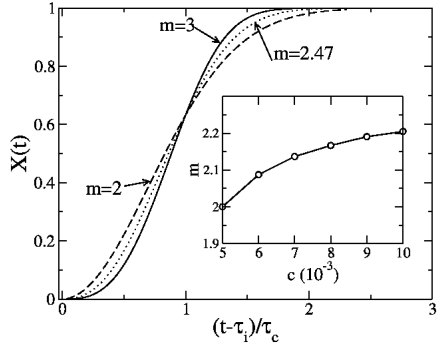} \end{center}
\caption{Fraction of transformed phase  $X(t)$ vs  scaled time for a 
$500\times 500$ system, with $W=1$, $\tau=0.25$, $\lambda=0.8$, 
$\Delta x=1$, $\Delta t=0.005$, averaged over $100$ ensemble averages 
using a concentration of impurities
$c=0.005$, and $k_BT=0.01$ (dashed line), $k_BT=0.027$ (dotted line), 
and $k_BT=0.05$ (solid line). Inset: KJMA exponent $m$ for different
values of the concentration of impurities $c$, with $W=1$, $\tau=0.25$, $\lambda=0.8$,
$\Delta x=1$, $\Delta t=0.005$, and $k_BT=0.01$.} 
\label{kjma_m_pf}
\end{figure}

As we have mentioned, the other interesting geometry is circular one. Thus 
Fig.\ \ref{impur_circ} 
shows how nucleation is enhanced at the boundary of a circular impurity. 
Moreover, we can use this result to relate the KJMA exponent, $m$, and the 
effect of impurities using a finite concentration $c$, of circular 
impurities of small radius. The results are plotted in Fig.\ \ref{kjma_m_pf}, 
where time has been rescaled to reveal the differences between
several $m$ values, for different temperatures. At this point, we want to 
stress that this dependence of $m$ on temperature has been obtained 
qualitatively in some Pd$_{1-x}$Si$_x$ crystallization experiments~\cite{Price}.
Furthermore, if we change the concentration of impurities, we can also modify the
value of $m$ (see inset in Fig.\ \ref{kjma_m_pf}). This result agrees  with
that in Ref.\ ~\cite{us2}. Thus the effect of temperature in KJMA equation is
not only present in the characteristic times, $\tau_i$ and $\tau_c$, but
even in the exponent $m$ in a nontrivial way. 

\section{Generalized heterogeneous nucleation}
\label{iv}

The results reported in the last section demonstrates that our phase field model is
a powerful tool to further advance in the knowledge and modeling of 
nucleation and growth phenomena,
in the homogeneous case and in the situation where impurities
catalyze the transformation in a subtle way.
Notwithstanding, the heterogeneities of a sample are not always due to
isolated  impurities. This is the main reason for the 
generalization that we introduce in this section.

We will assume that nucleation is heterogeneous, not in a phenomenological
way as in other proposed models~\cite{Frost}, but sticking to the classic 
ideas due to Cahn~\cite{Cahn}. Thus the system contains regions with 
some extra energy (for instance due to some order produced during deposition 
of the amorphous material), or 
at which nucleation is more probable. 
Let us show how we can cast this model on a mathematical footing.

\subsection{Quenched white noise}
\label{iva}

\begin{figure}[!ht]
\begin{center} \epsfxsize=9cm \epsffile{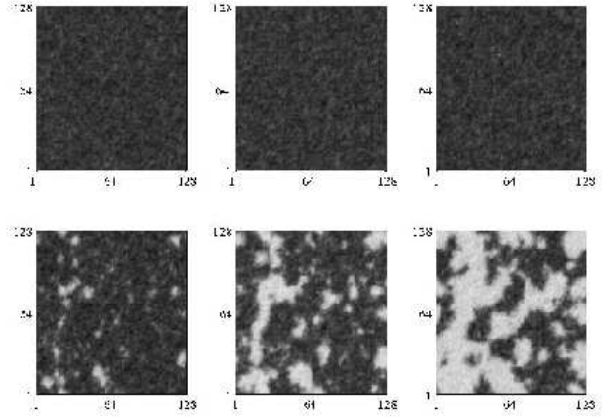} \end{center}
\caption{Comparison between homogeneous nucleation ($\Lambda=0$, top panels) 
and heterogeneous nucleation with quenched white noise ($\Lambda=0.3$, bottom 
panels), at different times (from left to right:  $t=20$, $50$, and $100$) with:
$W=1$, $\tau=1$, $dx=1$, $dt=0.005$, $k_BT=0.1$, and $\lambda=0.4$. }
\label{quenched_white}
\end{figure}

An intuitive way to introduce the latter idea in our model is sustained on
the assumption that the chemical potential depends locally
on position, 
\begin{equation}
\lambda\rightarrow\lambda+\rho({\bf r}),
\label{rho_def}
\end{equation}
$\rho({\bf r})$ being a Gaussian random variable
with zero mean and correlations given by
\begin{equation}
\langle\rho({\bf r})\rho({\bf r}^\prime)\rangle=\Lambda
\delta({\bf r} -{\bf r}^\prime),
\label{white_noise}
\end{equation}
where $\Lambda$ measures the degree of heterogeneity of the sample.
The main part of this generalization is that it does not make any specific 
assumption about the physical system under consideration. Unfortunately, 
numerical simulations suggest that it
cannot provide a KJMA exponent $m$, different from $3$ (homogeneous 
nucleation)~\cite{Karttunen}. Nevertheless, before exploring new
possibilities, we will show that the quenched white noise 
given by Eq.\ (\ref{white_noise}) unveils that disorder enhances 
nucleation (Fig.\ \ref{quenched_white}).

The fact that white noise is not capable to provide realistic values
of $m$ stems with intuition but, actually, it is very physical. Let
us make some calculations to show it clearly. 
If we perform again the thin interface limit of Eq.\ (\ref{pf}) with 
$\lambda\rightarrow\lambda+\rho({\bf r})$, we obtain the stochastic version of
Eq.\ (\ref{thin_interface}). Hence:
\begin{equation}
\Delta{\mathcal F}=2\pi R \sigma_l-\pi R^2\Delta \mu/\Omega+\chi(R),
\end{equation}
where $\chi(R)$ is given by 
\begin{equation}
\chi(R)=\delta g\int d{\bf r}\rho({\bf r}){\Theta}(r-R),
\end{equation}
$\Theta$ being the  Heaviside step function. The new noise term $\chi(R)$ 
has also zero mean and correlations given by:
\begin{equation}
\langle\chi(R)\chi(R^\prime)\rangle=\delta g\Lambda\pi R^2\delta(R-R^\prime).
\end{equation}
For brevity, we denote $F=\Delta{\mathcal F}$,  
$F_0=2\pi R \sigma_l-\pi R^2\Delta \mu/\Omega$ and  
$D(R)=\delta g\Lambda\pi R^2$.
Consequently, for every value of $R$ the free energy is a random number with
a distribution given by (recall that $\rho$ is Gaussian)
\begin{equation}
P(F)=\frac{1}{\sqrt{2\pi D(R)}}e^{-(F-F_0)^2/2D(R)}.
\end{equation}
Taking into account that $\exp[-\Delta{\mathcal F}(R)/k_BT]$ can be
understood as the barrier that has to be overcome to create a nucleus
of radius $R$, the critical radius will
be simply the one which minimizes that barrier. In the simplest case where
$\Lambda=0$, this minimum is the same as the maximum of $\Delta {\mathcal F}$,
but the situation changes because now $\Delta {\mathcal F}$ is a random 
variable. Thus in the physical system, we can only obtain information
about its mean value:
\begin{equation}
\langle e^{-F/k_BT}\rangle=\int dF e^{-F/k_BT}P(F)=\exp\left[\frac{D(R)}{2(k_BT)^2}-
\frac{F_0}{k_BT}\right]. 
\end{equation}
To be more specific, the critical radius is the solution of
\begin{equation}
\frac{d}{dR}\left(\frac{D(R)}{2k_BT}-F_0\right)_{R=R^*}=0\Rightarrow 
R^*=\frac{2\sqrt{2}W}{3\delta g(\lambda+\Lambda/2k_BT)}.
\label{r_critico_quenched}
\end{equation}
This is the result we were seeking: The critical radius diminishes
so that the nucleation rate increases and also depends on
temperature. But, as we anticipated above, the only effect
of quenched white noise is changing the time and length  scales of the
system, but not the
nucleation conditions necessary to obtain $m<3$. Moreover, if we
re-scale both length and time, then the systems with $\Lambda=0$ or 
$\Lambda\neq 0$ cannot be distinguished.
There is another interesting point of this theory which should be
remarked. In some experiments of recrystallization the {\em a priori}
value of $\Delta{\mathcal F}^*$ is actually much smaller than
that measured from the experiments~\cite{Doherty}, confirming that 
there is some kind of disorder which affects significantly this magnitude.

\subsection{Quenched colored noise}
\label{ivb}

\begin{figure}[!ht]
\begin{center} \epsfxsize=7cm \epsffile{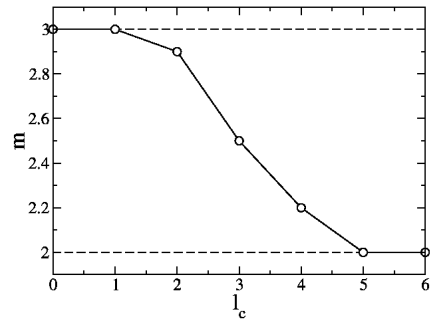} \end{center}
\caption{KJMA exponent, $m$, for different values of the correlation length,
using $W=1$, $\tau=0.15$, $\lambda=0.5$, $\Delta x=1$, $\Delta t=0.05$,
$k_BT=0.05$ and $\Lambda=0.05$.
($\circ$) and solid line, results from simulations. 
The dashed lines are a guide to the eye.} 
\label{Optimo_m_vs_l_c}
\end{figure}

At this point, we must collect some of the successful ingredients of
the theory and try to find out which must be the new ones in order
to provide a general model of heterogeneous nucleation. Then, the main question
is: What was implicit in the approach to heterogeneous nucleation
in Sec.\ \ref{iii} which is missing in Sec.\ \ref{iva}?
The answer is related to the new length scale introduced in Sec.\ \ref{iii} through 
the concentration  of impurities (and which is proportional to $c^{-1/2}$). 
Therefore we need both ingredients: disorder and 
a new independent length scale.
The simplest way to include both is by means of a quenched 
colored noise term:
\begin{equation}
\langle\rho({\bf r})\rho({\bf r}^\prime)\rangle=\Lambda
e^{-({\bf r}-{\bf r}^\prime)^2/2l_c^2}.
\label{gauss_corr}
\end{equation}
For the sake of completeness we have performed the simulations below
using Ornstein-Uhlenbeck correlations~\cite{Gardiner} instead of
Gaussian as in Eq.\ (\ref{gauss_corr}), but the results 
did not change significantly.
With this choice we have the same free-energy distribution of
probability $P(F)$ as in the white noise case, and also introduced
the required length scale.
As we expected, the KJMA exponent $m$ depends on the correlation length
$l_c$ as we show in Fig.\  \ref{Optimo_m_vs_l_c} as also  the
nucleation rate is larger than in the homogeneous case.
These results can be straightforwardly understood in terms of the involved 
characteristic length scales, the nucleation and growth length scale, 
$l=(V/N)^{1/3}$, and the quenched noise correlation length, $l_c$.
Thus if $l_c\ll l$ then the system
hardly see the defects, and it is almost homogeneous, so $m\simeq 3$. 
On the contrary, if $l_c\gg l$, as grains nucleate preferentially at the defects, and
they grow so fast that they impinge to each other at a mean distance $l_c$, so  
there is not enough time to allow other grains to nucleate.
Consequently, $m\simeq 2$ because every grain that nucleates does it mainly at the beginning
of the transformation. 
Note that these assumptions are valid whenever both $l$ and
$l_c$ are much smaller than the system size $L$. Notwithstanding, some care must be taken 
when simulating the model using large values of $l_c$ because the nucleation events 
are restricted to the surroundings of the defects so there are not many grains to 
sustain the validity of the KJMA equation.

Therefore the latter result contains relevant physical information relative to
the basic conditions yielding an exponent $m<3$. For instance,
it is consistent with those crystallization experiments of
Si$_{1-x}$Ge$_x$ by Olivares {\em et al.}~\cite{Olivares} or
recrystallization of  Pd$_{1-x}$Si$_x$ by Price~\cite{Price}. In the
first case that correlation length could be related to the concentration
of preexisting SiGe crystals created during deposition, and in the second 
case with the former polycrystalline structure which has been deformed before 
recrystallization.

\section{Discussion and Conclusions }
\label{v}

In this paper we investigated a phase field model 
modeling nucleation and growth events, such as crystallization,
island formation, or recrystallization. Starting from a Langevin
equation we have shown how it is a simple (and computationally efficient)
way to explore homogeneous nucleation, in terms of the well-known
description provided by the KJMA equation, Eq.\ (\ref{cryst_frac}).
Taking this equation as a starting point, we have provided 
several approaches to the problem of heterogeneous nucleation. The
first approach is related to the existence of defects which
are inert for the main growing material, such as wall domains or 
impurities. In both cases we have shown how nucleation is enhanced
near those defects. Moreover, a finite concentration
of isolated defects is able to modify the value of the KJMA exponent $m$,
which takes values lower than $m=3$, the known value for homogeneous
nucleation.  This kind of defects can
be suitable to study secondary crystallization, island nucleation in
heteroepitaxial systems (if one of the growing particles presents a slow 
dynamics compared to the others), or nucleation at step edges in terrace
growth. In particular, wall defects can be used to investigate
the influence of a terrace step on island formation (note that
$\lambda$ plays the role of the flux of particles arriving at the 
terrace). In fact, the competition
between dimer formation and nucleation at step edges has been reported
in Ref.\ \cite{Castellano}, and can be graphically seen in Fig.\ 
\ref{impur_pared}. Generalizing the model to higher dimensions, this kind
of wall defect could simulate the glass substrate on which the amorphous
material is grown prior to crystallization.
In the context of amorphous crystallization, circular defects can also be 
understood as crystals created at the growing stage, before crystallization 
takes place.

We have also presented a generalized model for heterogeneous nucleation
which deals statistically with imperfections or impurities by means
of a quenched noise term. The main result is related to the fact that this
quenched noise must be spatially correlated to provide a general
description in terms of the KJMA exponent.
As we mentioned above, the correlation length can be related to
preexisting crystalline regions embedded in the amorphous host or,
some deformed polycrystalline structure, but it may also be used to
quantify  the effect on island formation 
of mechanical stresses caused by the growth of multilayer devices.

Due to the general character of the model, it can be easily generalized.
For instance, a model that takes into account the possible
time dependence of the disorder should include two fields: the phase-field and
another one related to the motion of impurities. Besides this, sometimes the nucleating 
grains present some kind of anisotropy that should
be included in the free-energy functional, given by 
Eq.\ (\ref{Gibbs_funcional}).
Moreover, the model presented is only suitable for isothermal transformations so,
to study non isothermal effects, we should couple the phase field
with an evolution equation for the temperature, similar to some
prescriptions provided in the context of non-isothermal eutectic 
crystallization~\cite{Elder2}.  
Finally, there is some work left to do related to the suitable range
of parameters which allow comparisons with experimental data and also to
improve the efficiency of the numerical simulations, in particular when using
circular impurities to simulate heterogeneous nucleation.

\acknowledgements

The author thanks  A.\ S\'anchez, R.\ Gonz\'alez Cinca, and T.\ Arribas for 
useful discussions and helpful comments, and F.\ Dom\'{\i}nguez-Adame
for a critical reading of this manuscript.
This work has been partially supported by DGES (Spain) Grant No.\ BFM2000-0006.

\end{multicols}


\begin{references}

\bibitem[*]{ll} Email address: {\tt marioc@upco.es} 

\bibitem{Gunton} J.\ D.\ Gunton, J. Stat. Phys. {\bf 95}, 903 (1999).

\bibitem{Doherty} R.\ D.\ Doherty, Prog. Mater. Sci. {\bf 42}, 39 (1997).

\bibitem{Castellano} C.\ Castellano and P.\ Politi, Phys. Rev. Lett. {\bf 87}, 
	056102 (2001), and references therein.

\bibitem{Bergmann} R.\ Bergmann, G.\ Oswald, M.\ Albrecht, and J.\ H.\ Werner,
	Solid State Phenom. {\bf 51-52}, 515 (1996).

\bibitem{Im} J.\ S.\ Im and R.\ S.\ Sposili, MRS bull. {\bf 21}, 
	39 (1996).

\bibitem{Uemoto} Y.\ Uemoto, E.\ Fujii, A.\ Nakamura, K.\ Senda, and H.\
Takgi, IEEE Trans. Electron Devices {\bf ED-39}, 2359 (1992).

\bibitem{Meso} Hereafter we will refer to impurity as a {\em mesoscopic} inert 
	object for the growing material, to be understood as strange particle or
	imperfection.

\bibitem{Olivares} J.\ Olivares, A.\ Rodr\'{\i}guez, J.\ Sangrador, T.\ 
	Rodr\'{\i}guez,  C.\ Ballesteros, and A.\ Kling, Thin Solid Films {\bf
	337}, 51 (1999).

\bibitem{us1} M.\ Castro, A.\ S\'anchez, F.\ Dom\'{\i}nguez-Adame, and 
	T.\ Rodr\'{\i}guez, Appl. Phys. Lett. {\bf 75}, 2207 (1999). 

\bibitem{us2} M.\ Castro, A.\ S\'anchez, and  F.\ Dom\'{\i}nguez-Adame, 
	Phys. Rev. B {\bf 61}, 6579 (2000). 

\bibitem{Grant} M.\ Grant and J.\ D.\ Gunton, Phys. Rev. B {\bf 32}, 
	7299 (1985). 

\bibitem{Roy} A.\ Roy, J.\ M.\ Rickman, J.\ D.\ Gunton, and K.\ R.\ Elder,
	Phys. Rev. E {\bf 57}, 2610 (1998).

\bibitem{Rikvold} R.\ A.\ Ramos, P.\ A.\ Rikvold, and M.\ A.\ Novotny,
	Phys. Rev. B {\bf 59}, 9053 (1999).

\bibitem{Jou} H.-J.\ Jou and M.\ T.\ Lusk, Phys. Rev. B {\bf 55}, 8114 (1997). 

\bibitem{Kolmogorov} A.\ E.\ Kolmogorov, Bull. Acad.\ Nauk.\ USSR., 
	Mat.\ Ser.\ {\bf 1}, 355 (1937); W.\ A.\ Johnson and R.\ F.\ Mehl, 
	Trans.\ AIME\ {\bf 135}, 416 (1939); 
	M.\ Avrami, J.\ Chem.\ Phys.\ {\bf 7}, 103 (1939). 

\bibitem{Karpov} V.\ G.\ Karpov, Phys. Rev. B {\bf 50}, 9124 (1994);
	V.\ G.\ Karpov and D.\ W.\ Oxtoby, {\em ibid.} {\bf 54}, 9734 (1996).

\bibitem{Liu} X.\ Y.\ Liu, J. Chem. Phys. {\bf 112}, 9949 (2000).

\bibitem{Enomoto} Y.\ Enomoto, Acta Metall. Mater. {\bf 38}, 173 (1990).

\bibitem{Weinberg} M.\ C.\ Weinberg, J. Non-Cryst. Solids {\bf 225}, 1 (1999).


\bibitem{Kobayashi} R.\ Kobayashi, Physica D {\bf 63}, 410 (1993).

\bibitem{Wheeler} A.\ A.\ Wheeler, B.\ T.\ Murray, and R.\ J.\ Schafer, 
	Physica D {\bf 66}, 243 (1993).

\bibitem{Karma} A.\ Karma and W.-J.\ Rappel, Phys. Rev. E {\bf 57}, 
	4323 (1998).

\bibitem{Folch} R.\ Folch, J.\ Casademunt, A.\ Hern\'andez-Machado, 
	and L.\ Ram\'{\i}rez-Piscina, Phys. Rev. E {\bf 60}, 1724 (1999); 
	{\bf 60}, 1734 (1999).

\bibitem{Elder} K.\ R.\ Elder, M.\ Grant, N.\ Provatas, and J.\ M.\
	Kosterlitz, Phys. Rev. E {\bf 64}, 021604 (2001).

\bibitem{Collins} J.\ B.\ Collins and H.\ Levine, Phys. Rev. B 
	{\bf 31}, 6119 (1985).

\bibitem{Tapio} M.\ Dub\'e, M.\ Rost, K.\ R.\ Elder, M.\ Alava, S.\	
	Majaniemi, and T.\ Ala-Nissila, Eur. Phys. J. B {\bf 15}, 701 (2000).

\bibitem{Steinbach} I.\ Steinbach, F.\ Pezolla, B.\ Nestler, J.\ Rezende,
	M.\ Seesselberg, and G.\ J.\ Schmitz, Physica D {\bf 94}, 135 (1996).

\bibitem{Nestler} B.\ Nestler and A.\ A.\ Wheeler, Phys. Rev. E {\bf 57},
	2602 (1998).
	
\bibitem{Kobayashi2} R.\ Kobayashi, J.\ A.\ Warren, and W.\ C.\ Carter, 
	Physica D {\bf 140}, 141 (2000);  J. Cryst. 
	Growth {\bf 211}, 18 (2000).

\bibitem{Lobkovsky} A.\ E.\ Lobkovsky and J.\ A.\ Warren, Phys. Rev. E 
	{\bf 63}, 051605 (2001).

\bibitem{phi0} The level curve $\phi=0$ is generally accepted to define 
	the boundary position. Notwithstanding, this is not necessary and, in 
	order to obtain the correct sharp-interface limit of some phase-field models,
	other values can be more suitable (see Ref.\ \cite{Elder} for further
	details).

\bibitem{Hohenberg} P.\ C.\ Hohenberg and B.\ I.\ Halperin, Rev. Mod. Phys. 
	{\bf 49}, 435 (1977).

\bibitem{other} For completeness we have reproduced the numerical 
	simulations presented in Secs. \ref{iii} and \ref{iv} using 
	$g(\phi)=\phi-2\phi^3/3+\phi^5/5$,
	but the results were not affected by the specific choice of $g$.


\bibitem{cnt} R.\ Becker and W.\ D\"oring, Ann. Phys. (Leipzig) {\bf 24},
	719 (1935); Ya.\ B.\ Zeldovich, Acta Physicochim. (URSS) {\bf 18}, 1 (1943).
	
\bibitem{Gardiner} C.\ W.\ Gardiner, {\em Handbook of Stochastic
	Methods} (Springer-Verlag, Berlin, Heidelberg, 1985).

\bibitem{Elder_eu} K.\ R.\ Elder, F.\ Drolet, J.\ M.\ Kosterlitz, and M.\ 
	Grant, Phys. Rev. Lett. {\bf 72}, 677  (1994).

\bibitem{Karttunen} Similar results were found by M.\ Karttunen, N.\ Provatas, T.\ Ala-Nissila, 
	and M.\ Gardner, J. Stat. Phys {\bf 90}, 1401 (1998), in the context of flame fronts
	in slow combustion.

\bibitem{Price} C.\ W.\ Price, Acta Metall.\ Mater.\ {\bf 38}, 727 (1990), 
        and references therein.


\bibitem{Frost} H.\ J.\ Frost and C.\ V.\ Thompson, Acta Metall.\ 
        {\bf 35}, 529 (1987).

\bibitem{Cahn} R.\ W.\ Cahn, J. Inst. Met. {\bf 76}, 121 (1949).

\bibitem{Elder2} K.\ R.\ Elder, J.\ D.\ Gunton, and M.\ Grant, Phys. Rev. E
	{\bf 54}, 6476 (1996).

\end{references}
\end{document}